\documentclass[10.5pt,compsoc]{report}
\usepackage{CJKutf8}
\usepackage{graphicx}
\usepackage{footmisc}
\usepackage{subfigure}
\usepackage{url}
\usepackage{multirow}
\usepackage[noadjust]{cite}
\usepackage{amsmath,amsthm}
\usepackage{amssymb,amsfonts}
\usepackage{booktabs}
\usepackage{color}
\usepackage{ccaption}
\usepackage{booktabs}
\usepackage{float}
\usepackage{fancyhdr}
\usepackage{caption}
\usepackage{xcolor,stfloats}
\usepackage{comment}
\graphicspath{{figures/}}
\usepackage{cuted}  %flushend,
\usepackage{epstopdf}
\UseRawInputEncoding
\usepackage{mathtools, nccmath}

\usepackage{listings}
\usepackage{color}
\definecolor{dkgreen}{rgb}{0,0.6,0}
\definecolor{gray}{rgb}{0.5,0.5,0.5}
\definecolor{mauve}{rgb}{0.58,0,0.82}
\begin{document}

\thispagestyle{empty}

\hyphenpenalty=50000

\makeatletter
\newcommand\mysmall{\@setfontsize\mysmall{7}{9.5}}
\renewcommand\thesection{\arabic{section}}
%%%%%%%%%%%%%%%%%%%%%%%
\newenvironment{tablehere}
  {\def\@captype{table}}
  {}
\newenvironment{figurehere}
  {\def\@captype{figure}}
  {}
%%%%%%%%%%%%%%%%%%%%%%%
%%%%%%%%%%%%%%%%%%%%%%%%%%%%%%%%

\thispagestyle{plain}%
\thispagestyle{empty}%

\let\temp\footnote
{}
\begin{center}
{\large\textbf{
Internet Of Rights(IOR) In Role Based Block Chain
}}
\end{center}

\begin{center}
{\sf 
Yunling Shi, Jie Guan, Junfeng Xiao, Huai Zhang, Qiang Guo 1， Yu Yuan 2
}
\end{center}

\vspace{3.5mm}
{\noindent
\bf{Abstract:} {\sf
A large amount of data has been accumulated. with the development of the Internet industry. Many problems have been exposed with data explosion: 1. The contradiction between data privacy and data collaborations; 2. The contradiction between data ownership and the right of data usage; 3. The legality of data collection and data usage; 4. The relationship between the governance of data and the governance of rules; 5. Traceability of evidence chain. In order to face such a complicated situation, many algorithms were proposed and developed. This
article tries to build a model from the perspective of blockchain to make some breakthroughs.Internet Of Rights(IOR) model uses multi-chain technology to logically break down the consensus mechanism into layers, including storage consensus, permission consensus, role consensus, transaction consensus etc., thus to build a new infrastructure, which enables data sources with complex organizational structures and interactions to collaborate smoothly on the premise of protecting data privacy. With blockchain’s nature of decentralization, openness, autonomy, immutability, and controllable anonymity, Internet Of Rights(IOR) model registers the ownership of data, enables applications to build ecosystem based on responsibilities and rights. It also provides cross-domain processing with privacy protection, as well as the separation of data governance and rule governance. With the processing capabilities of artificial intelligence and big data technology, as well as the ubiquitous data collection capabilities of the Internet of Things, Internet Of Rights(IOR) model may provide a new infrastructure concept for realizing swarm intelligence and building a new paradigm of the Internet, i.e. intelligent governance.}

\vspace{3.5mm}
{\noindent

\thispagestyle{plain}%
\thispagestyle{empty}%
\makeatother

\begin{figure}[b]
\vskip -6mm
\begin{tabular}{p{44mm}}
\toprule\\
\end{tabular}
\vskip -4.5mm
\noindent
1.Beijing QingZhi ShuYuan Technology Co., Ltd; 

2.IEEE
\setlength{\tabcolsep}{1pt}
\begin{tabular}{p{1.5mm}p{79.5mm}}
%{$\sf{\hskip 0.5em*}$}

\end{tabular}
\end{figure}\large

\clearpage

\vspace{3.5mm}
\section{Introduction}
\label{s:introduction}
\noindent
In the data era, information technology can help people better perceive surroundings, build smooth communication channels, and assist decision-making. Industries such as the Internet of Things, cloud computing, big data, and the Internet are generating data all the time, and the accumulation of data accelerates. In order to provide better data services for applications such as artificial intelligence and big data, the ownership, legality, and privacy of data resources, as well as the relationship between rule makers and regulators, rules and data, need to be sorted out. Therefore, the Internet of Rights (IOR) came into being.

\vskip 6mm
\noindent
Considering the loose correlations and contradictory of data, Internet of Rights(IOR) model attempts to set up a new data sharing mechanism based on block chain, and build a swarm intelligence ecosystem that incorporates the Internet of Things and cloud computing. It has multiple features such as high efficiency, fairness, transparency, privacy protection, right protection, and supervision, therefore provides a solid data foundation for artificial intelligence and big data.

\section{Background}
\label{s:Background}

\subsection{Data privacy and collaboration contradiction}
\noindent
High-quality data is the foundation of artificial intelligence and big data. Data collaboration requires high quality and as much data as possible to provide support for artificial intelligence and big data. However, the owners of the data do not want data to be misused, especially the data that needs privacy protection. Therefore, they only want to provide as little data as possible. This is the contradiction between data privacy and data collaboration.

\subsection{Data legality}
\noindent
Under vague terms of user agreements, user data is often over-collected or used without notice, so the collectors will have legal risks. Collectors often do not have the capabilities of artificial intelligence and big data processing, and they need to entrust a third party to process data and do the calculations. This requires third parties and even more participants to have the right to legally use data, which introduces more complexity and security requirements to the system. On the other hand, it is usually difficult for artificial intelligence and big data to obtain legal data, which leads to slow development of such technologies.

\subsection{Conflict between ownership and the rights of usage}
\noindent
The “free of charge” mode in Internet is based on a centralized system, where the user data in the system is used for free. More and more users gradually realize the value of their own data, and require a declaration of data ownership, or even monetization. The use of data will gradually change from free mode to payment mode.

\vskip 6mm
\noindent
Currently, data is stored in a centralized way on the Internet. A large amount of data is averaged before it reaches date consumers, and the public has fewer and fewer opportunities to obtain raw data. Therefore, the value of personalized date, is obscured. Data is driven by data center instead of the requirements of data producers and data consumers. With the development of data, more users are willing to discover the value of data on a paid basis, so they need a new infrastructure to make this possible.

\subsection{Separation of data governance and rule governance}
\noindent
Since data privacy is often violated and data confirmation cannot be guaranteed, the common method is to centralize the data on a trusted management platform. This may cause the management system to be too large to manage. The responsibility with such system may also become a large burden. On the other hand, it caused the data to be used in a daunting manner and the value of data could not be efficiently discovered. Therefore, the data responsibilities, rights, and benefits need to be consistent. The governance of data, which refers to the behaviors of data owners and data users; and the governance of rules, which refers to the behaviors of rule makers and data flow regulators; are two different dimensions. The governance of data and governance of rules, need to be separated logically.

\vskip 6mm
\noindent
The trusted management platform is usually a rule maker or regulator. The aggregation of data to a trusted party breaks the logical relationship between data and rules. How to ensure that data and rules operate separately, and data flows effectively within the infrastructure is extremely important.

\subsection{The traceability of the evidence chain}
\noindent
The rule itself and the processes of rule execution need to be auditable for a period of time and can be used as evidence to verify that these have indeed happened. Therefore, the evidence should be traceable and immutable. A mechanism should be provided to demonstrate how the data and rules are created and updated. The original data and rules cannot be modified on logs, and can only be updated with explanations and signatures if necessary.

\subsection{The cost of preventing infringement}
\noindent
The common solution to prevent infringement is to detect infringement through comparison by human beings or by artificial intelligence. Infringement would be punished using legal means. In many scenarios, the cost of infringement detection and reduction is higher than the compensation itself from the lawsuit. Therefore, infringement is dealt case by case, and cannot be banned totally. 

\vskip 6mm
\noindent
Ideally, in addition to legal protection, mechanisms should also be introduced to make the legal usage cheaper than infringement, and the benefits of legal usage higher infringement.

\section{Overview}
\noindent
As a decentralized infrastructure provides a feasible methodology for the five issues mentioned above. Next, we will describe how Internet of Rights(IOR) builds a model of intelligent governance.

\subsection{Internet of Rights(IOR) model}
\noindent
In a centralized system, each center communicates via interfaces. This manner of data usage is not conducive to user communication, privacy protection, data sharing, and collaborations. Under the Internet of Rights(IOR) model as a decentralized infrastructure supports controllable anonymity, data sharing, collaboration, supervision, and traceability. It is a trusted bridge to connect data, and to connect data consumers with data producers, thus optimizes the allocation of resources.

\begin{center}
\includegraphics[width=0.9\textwidth]{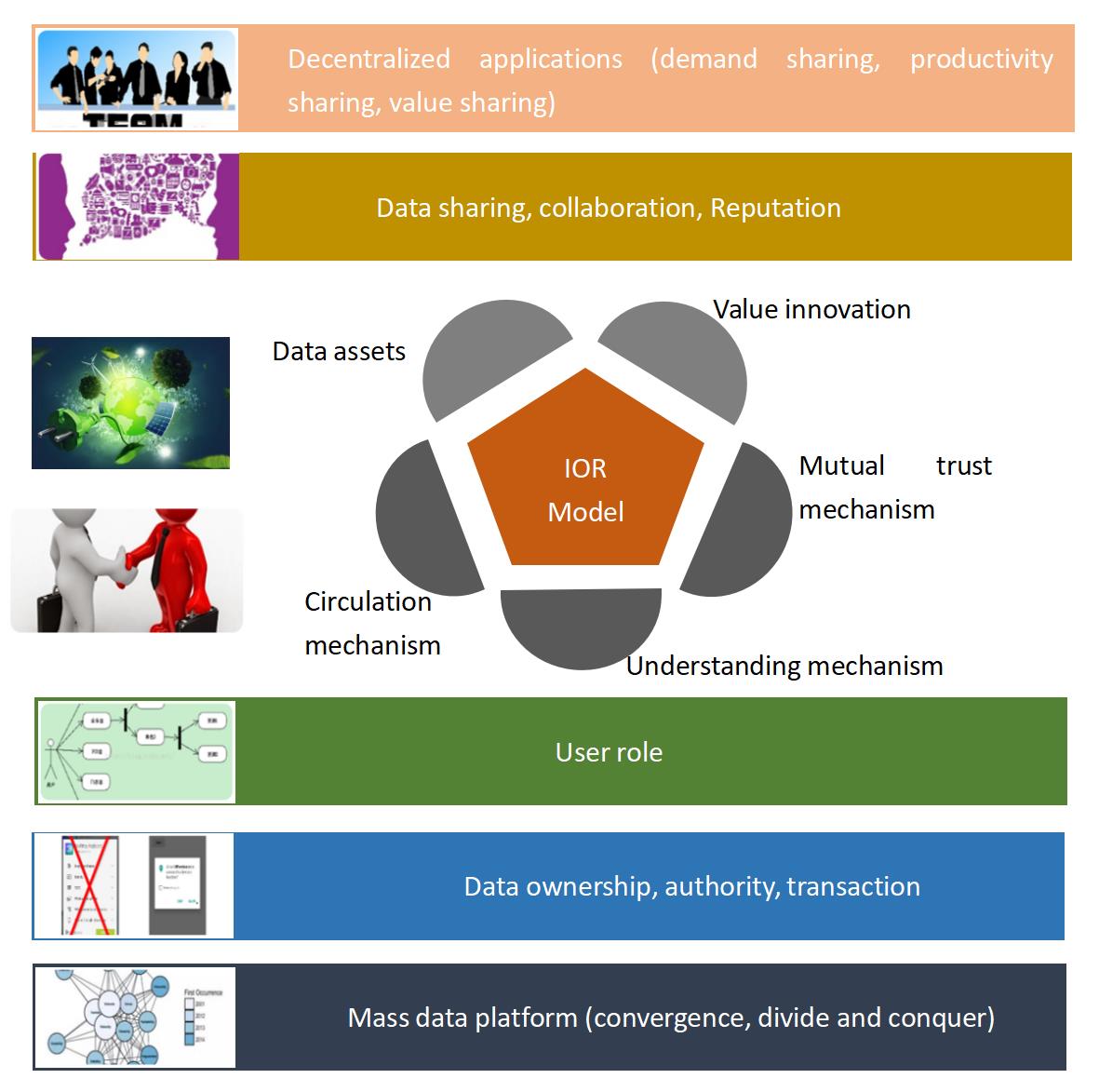}
\noindent
\begin{center}
\end{center}
{\normalsize\textbf{Fig. 1\quad Internet of Rights(IOR) model}}
\end{center}

\vskip 6mm
\noindent
By means of converging data assets, Internet of Rights(IOR) model establishing mutual trust and understanding mechanisms. Internet of Rights(IOR) model introduces roles and permissions into traditional block chain, and standardizing data governance and rule governance, in order to make legal usage of data throughout its life cycle. With identity management within blockchain, self-sovereign identity would be made possible, and a safe and efficient identity authentication and authority management mechanism would be created [13].

\subsection{Topology}
\noindent
Internet of Rights(IOR) model supports the characteristics of multi-chain and chain separation. It evolves from a flat structure to a tree structure. The main chain and sub-chains are logically separated chains that can be expanded and upgraded independently. Each sub-chain is connected to the upper chain through a group of leader nodes, which can speed up transaction submissions and ensure credibility [6].

\vskip 6mm
\noindent
Multi-chain is divided into three types: role chain, data access control chain or permission chain, and service chain. The service chain can be separated recursively, and the separated sub-chain contains at least a role chain and a permission chain. The fewer chain nodes submit transactions, the faster the consensus can be achieved. Therefore, the number of consensus nodes in the chain does not need to be large. In order to increase the reliability, we suggest that the nodes participating in the consensus be trusted nodes, and the rollup [7] process can be selected after the transaction is submitted, with minimum amount of data for the transaction proof is submitted to the upper chain [8].

\begin{center}
\includegraphics[width=0.9\textwidth]{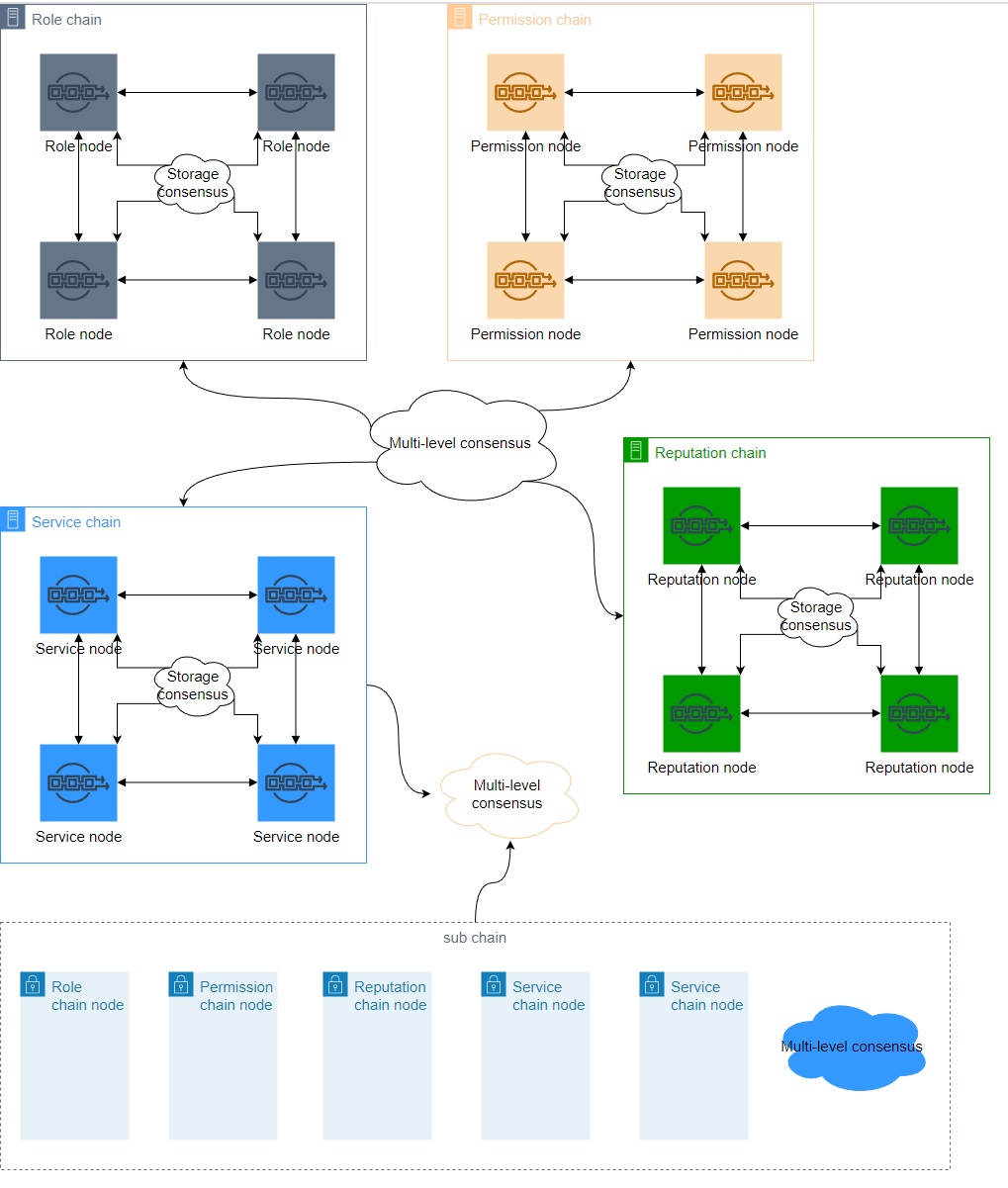}
\noindent
\end{center}
\begin{center}
{\normalsize\textbf{Fig. 2\quad Multi-level consensus}}
\end{center}

\vskip 6mm
\noindent
Big data technology focuses on massive data processing, storage, and calculation, while blockchain technology focuses on decentralization, immutability, controllable anonymity, and cross domain processing of information. They are perceived as two complementary strategies. Mass data should be processed by the big data system, but the fingerprints and metadata of permissions of the data are on the chain. The role chain and permission chain are the standardization of data governance rules, allowing an effective integration of blockchain technology and big data technology. The integration of blockchain technology with distributed file systems, big data analysis, cloud computing, artificial intelligence and other technologies is crucial for future development [14].

\subsection{Role}
\noindent
The role refers to the nature of the user, which classifies users with similar permissions. Management of all users and data by a centralized system will inevitably lead to data concentration, and unnecessary responsibility. The role clearly defines the responsibility boundaries of various participants, so that everyone can obtain the necessary data in compliance with rules and regulations. Participant can only access and operate data within the scope of their roles. The role definition is defined by the user manager and the metadata is submitted to the role center. The user manager assigns role identifiers on the role chain to ensure uniqueness in the role chain. The role chain does not provide role application services, but only role consensus services, that is, role data on the chain.

\vskip 6mm
\noindent
Role data mainly includes the relationship between users and roles. Not all role relationships of a user are stored in one role data structure, and one role data structure only contains one user. The modification of role data must be done through blockchain transactions.

\subsection{Permission}
\noindent
Permission refers to the collection of data access and use capabilities, which has subordinate permissions and is a tree structure. Permissions refer to dimensions of data categorizing and the ways of using data. Permission is an attribute of data and does not change with the changes of users.

\vskip 6mm
\noindent
Permissions and roles are separated. The authority center is only responsible for the definition of data usage attributes, not about who might have the permissions. The permission definition is defined by the regulators and metadata is submitted to the permission center. The regulators assign the authority identifiers on the authority chain to ensure that they are unique within the chain. The permission chain does not provide permission application services, but only permission consensus services.

\vskip 6mm
\noindent
Permission data includes two types of data: role and permission mapping, and data permission. The mapping of roles and permissions defines which permissions a role has. The permissions contained in a role are not necessarily contained in only one permission data structure, and one permission data structure only contains the permissions owned by one role. A data permission only includes all permission definitions for a data. The modification of data permissions must be done through blockchain transactions.

\subsection{Traceability}
\noindent
One of the basic capabilities of the blockchain based on UTXO technology is traceability, which increases the cost of fraud and maintains a trusted collaboration environment. The traceability of UTXO is realized by three features:

\vskip 6mm
\noindent
1. The vins field contains the hash value of the pre-UTXO (hash) and the owner's signature.

\vskip 6mm
\noindent
2. The vouts field contains the allocation of subsequent UTXOs, and the sum of their values is equal to the sum of the values of all preceding UTXOs. The hash value (hash) of the subsequent UTXO is represented by the hash value (hash) of the entire tx. 

\vskip 6mm
\noindent
3. UTXO cannot be double spent. It can only be spent once.

\vskip 6mm
\noindent
Therefore, the UTXO technology makes the source paths of all UTXO traceable through the vins and vouts fields, and has the corresponding UTXO owner's signature to ensure the legality of UTXO usage. UTXO also guarantees that there can only be one source path in the life cycle of a UTXO by prohibiting double spending.

\section{Multi-level consensus}
\label{s:Multi-level consensus}
\noindent
The introduction of smart contracts has brought many opportunities to the ecology of the blockchain, and the blockchain's support for data has made a big step forward. However, simply assigning all data logic to smart contracts will undoubtedly increase the complexity, and decrease the robustness of smart contracts. We use multi-level consensus to extract non-business logic in smart contracts, thus manage data and rules more effectively and securely. 

\vskip 6mm
\noindent
Multi-level consensus is bottom-up and has good scalability, including but not limited to: storage consensus, role consensus, and permission consensus. The scope of multi-level consensus will be broadened in the future.

\subsection{Storage consensus}
\noindent
The basis of multi-level consensus is storage consensus. Participants in different scenarios are different, so each scenario has an independent chain to provide consensus services. The sharing of information between multiple chains requires mutual authentications between the chains, i.e. the storage consensus. 

\vskip 6mm
\noindent
Storage consensus requires the following capabilities: 

\vskip 6mm
\noindent
1. Multi-chain addressing capability: Since the storage between two chains is not shared, a communication and addressing mechanism must be provided so that the chain can understand each other's location. 

\vskip 6mm
\noindent
2. Distributed storage capacity: The storage consensus itself relies on distributed storage to allow the data stored by each participant to reach agreement.

\vskip 6mm
\noindent
3. Distributed verification capability: When receiving a verification request that is not signed by the original chain, the authenticated data needs to be found and sent to the source chain for verification, based on the identification information of the source chain contained in the data itself, using addressing capability.

\vskip 6mm
\noindent
4. Distributed transaction capability: The transaction in a multi-chain system is distributed, and the participants of the transaction are not limited to one chain. In the UTXO transaction scenario, all input UTXOs are recovered by their respective source chains, and the recycled UTXOs will generate a corresponding new UTXO in the transaction chain. All output UTXOs are generated from these new UTXOs. Both recycled and newly generated UTXO are encrypted and signed by the private key designated by the system.

\vskip 6mm
\noindent
5. Distributed security capability: All communication channels need to be encrypted, and the identities of the participating parties are mutually verified. The security of the verification process is solved by cryptography, the communication channel uses ssl, and the identity verification uses an asymmetric encryption algorithm for abstract and signature.

\subsection{Role consensus}
\noindent
In a distributed decentralized network, data ownership belongs to users. The user saves the fingerprint of the role information through the storage consensus in the role chain.

\vskip 6mm
\noindent
The role data is accessed through the blockchain network as the signaling network communication. After the signaling network handshake, the private information is exchanged through the temporary trusted channel, and the access log is stored through hash collision. When needed in the future, it can be compared and verified on the blockchain network to ensure the authenticity of information exchange.

\vskip 6mm
\noindent
Role consensus is based on storage consensus. Role data can be stored on any role chain, be authenticated on any role chain, and participate in distributed transactions of any role chain.

\subsection{Permission consensus}
\noindent
The nature of multiple roles of producers and consumers determines that their permissions are also multidimensional. Among the data in each domain, there is a need for confidential data to be shared with other domains and subject to supervision. Therefore, the data provided by each domain must be honest and reliable, and there must be no data conflicts on different occasions. However, it does not want to be publicly visible, nor does it want the administrator to have excessive authority.

\vskip 6mm
\noindent
The fingerprint of the permission data is uploaded to the chain to form a permission consensus. According to the multidimensional authority standard. The definition and verification of distributed authority are realized under the framework of distributed roles, and data assembly and packaging capabilities are provided. Permission data is transmitted through a temporary encrypted channel and is not stored on the blockchain network, preventing immutable information from being cracked by more advanced cryptographic schemes in the future, avoiding data leakage, and protecting user privacy.

\vskip 6mm
\noindent
Permission consensus is based on storage consensus. Permission data can be stored in any permission chain, authenticated on any permission chain, and participate in distributed transactions in any permission chain.

\subsection{Reputation consensus}
\noindent
Reputation refers to the influence of an entity calculated based on user data, user behavior, intellectual property and digital assets. Reputation includes but not limited to reward points, and credits. Reputation reflects the entity’s influence, which can be extended to its derivatives and interaction with other entities.

\vskip 6mm
\noindent
Reputation consensus refers to the consensus of influence of a subject reached within a scope and is recorded in blockchain. Reputation generally agreed on blockchain would be a better replacement of page ranking for searching engines. A good reputation will gain high visibility and will gain more potential benefits, while a poor reputation will reduce potential benefits. Through reputation consensus, the subject will be encouraged to act in the manner promoted by the rules of consensus within its scope.

\section{Protocol}
\label{s:Protocol}

\subsection{Protocol data structure}
\noindent
In Internet of Rights(IOR) model, blockchain is based on the UTXO extended model [2], uses the dyeing model as the domain division, and also adds the ability to verify historical scenes. The transaction data structure is divided into several parts: vins, vouts, color, reference, cooperation, tx data, tx signature, block data. These designs ensure that all data is verifiable, and responsibilities, rights, and benefits are bounded.

\vskip 6mm
\noindent
The “color” field represents the color model, and different colors represent different scenes. The “reference” field indicates the entity association and can be used for entity indexing. The “cooperation” field is a multi-party collaboration signature. The “tx” data field is transaction data. The “tx” signature field is the transaction data signature field, and the tx data field is protected from modification. Block data contains the block number and transaction number, which are used for historical scene verifications. 

\vskip 6mm
\noindent
Due to the multi-chain technology, the number of consensus nodes in each chain is reduced, and 51\% attacks would become easier. The GHOST[5] algorithm puts forward the principle of maximizing the number of blocks after the fork, not only the principle of the longest block of the fork, which increases the difficulty of 51\% attacks. The Conflux [9] algorithm also adopts the principle of maximizing the number of bifurcation blocks, combined with directed acyclic graph (DAG) calculations to improve performance and security. The intelligent governance model multi-chain technology, may use the advantages of these two technologies, combined with trusted node verification and the Rollup mechanism [7], to further enhance performance and security. The attacker has to forge the longest chain and the number of nodes [5], which would only break through the protection of the first layer of chain. It also needs to break through the second layer of the Rollup chain to make its forgery take effect, so the degree of difficulty is doubled.

\subsection{Authorization process}
\noindent
The owner A of entity E manages the authority of entity E in the blockchain system, and uses UTXO U1 as the proof of ownership, where the value of U1 is the number of authorizations. User B of entity E needs to apply for permission from A before using it. A converts his U1 into two new UTXOs through the UTXO model. One is to provide B with UTXO U2 for the usage of E, and the other is A's ownership UTXO U3. The sum of the value of U2 and U3 is equal to the value of U1.

\vskip 6mm
\noindent
Similarly, when user B uses multiple entities to generate a new entity F, he will obtain multiple UTXOs in turn. When user B publishes a new entity, he converts these multiple UTXOs into the ownership UTXO of a new entity. The conversion of the right of use UTXO to the ownership UTXO needs to be signed by multiple parties, including: the process party, the right to use UTXO party, and the owner of the UTXO party. 

\vskip 6mm
\noindent
When user C uses entity F, he needs to apply for authorization from owner B, but user C does not need to apply for authorization from owner A. Since user B has applied for authorization from owner A, the traceability of the authority is guaranteed. 

\vskip 6mm
\noindent
In this process, each entity's permission UTXO (including usage rights and ownership) is equivalent to a point, and each authorization is equivalent to an edge, which forms a directed graph, and all authorization models can be calculated through graph calculations. The value of UTXO represents the number of permissions and is separable. It is modified every time it is converted, and the life cycle management of the permission can be carried out. The directed graph records all compliance and permission transfer information, and its relationship can be calculated based on the graph to get a reasonable result, ensuring the legitimate rights and interests of owners and users.

\begin{center}
\includegraphics[width=0.9\textwidth]{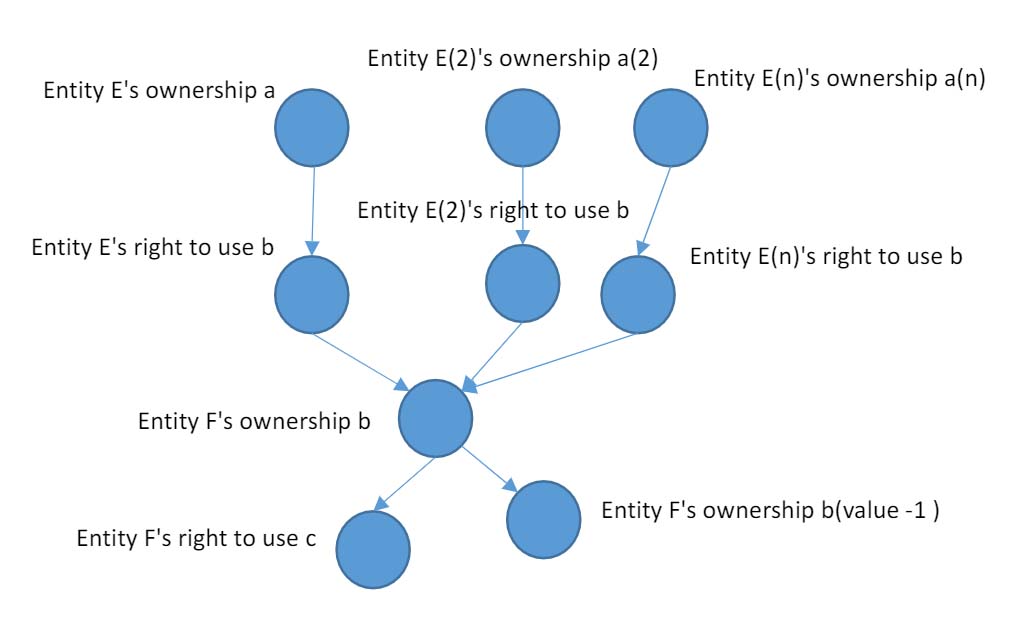}
\noindent
\end{center}
\begin{center}
{\normalsize\textbf{Fig. 3\quad Authorization process}}
\end{center}

\subsection{Multi-party collaborative signature}
\noindent
The multi-party cooperative signature adopts the proxy signature [3] mechanism, and requires the orderliness of the cooperative signature. First, the reference field stores the hash value (Hash) of the multi-party collaboration process UTXO, and the UTXO can verify the authorization of the participants and the collaboration sequence of the multi-party collaboration. Second, each participant provides its own UTXO and its signature. Finally, the transaction committer signs the entire transaction data, saves it in the tx signature field, and submits it. Multi-party collaborative signatures support each step of signing data on the chain, but the intermediate process cannot be authenticated and is only stored as a log.

\vskip 6mm
\noindent
The signature algorithm adopts verifiable random function $VRF(x,sk) $ algorithm[1], input string $(x)$ and private key $(sk)$, VRF algorithm returns hash value $ (hash) $ and proof the value is $ (proof) $. The proof value ($ proof $, denoted by $ \pi $) can be calculated from the public key ($ pk $) and hash value ($ hash $) for verification. Since the private key ($ sk $) is only owned by the user, it can prove the validity of the signature.

\begin{center}
\includegraphics[width=0.9\textwidth]{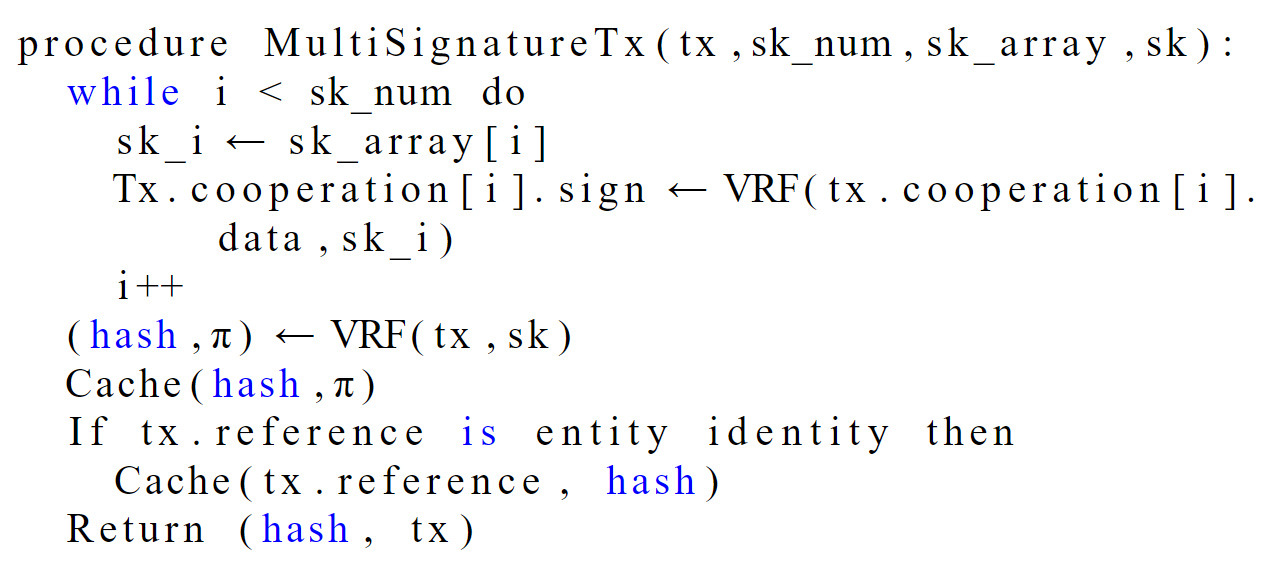}
\noindent
\end{center}
\begin{center}
{\normalsize\textbf{Fig. 4\quad Signature algorithm function}}
\end{center}

\vskip 6mm
\noindent
Since the cooperation will add a hash of the participant each time the multi-party collaborative signature is transferred, the cooperation part is only used to prove that the participant owns the previous data. When data is on the chain, the blockchain consensus node fills in the current block number and its own signature for historical validity verification.

\subsection{Data authentication}
\noindent
Transaction data (TX) can be authenticated through the inverse operation process of multi-party collaborative signatures, and is divided into two types of authentications:

\vskip 6mm
\noindent
1. Current validity authentication. 

\vskip 6mm
\noindent
2. Historical validity verification.

\vskip 6mm
\noindent
Log data is all historical data in the blockchain system, and historical validity can be used to verify whether it was legal. The valid history does not indicate the validity as of now, and the validity as of now indicates a valid history. 

\vskip 6mm
\noindent
Validity check of current data: First, find the transaction data itself on the chain according to the transaction hash value from the blockchain, check whether it is unspent, and then verify whether all the multi-party cooperative signatures of the transaction data are legal and valid.

\vskip 6mm
\noindent
Validity check of historical data: Find the corresponding block through the block number in the transaction data to be verified, then find the corresponding transaction data from the block, check whether the transaction to be verified is included, and finally verify all the multi-party cooperation of the transaction data in turn to see if the signature is legal and valid.

\subsection{Rules of protocols}
\noindent
The data is effectively segmented through the role and permission system, and different participants use different data according to the role definition, so that the necessary data is available when the data is needed. Data beyond the scope of authority will not be accessible, which fully protects data privacy. Users no longer worry about excess data being used at will, and the promised data usage range will not be exceeded.

\vskip 6mm
\noindent
All the fingerprints of role and permission data need to be chained, and the collected behavior data needs to be chained after data collection. Collecting behavioral data refers to: who, when, how, and the signature of the data. This mechanism is adopted to ensure the privacy is not violated. Evidence is provided to check if the data is used in compliance and lawfulness.

\vskip 6mm
\noindent
When data, roles, and permissions are on the chain, they must be signed by multiple parties. The application process and results of roles and permissions are all signed. These signatures are the authentication of the ownership and use rights of the data. If a new right to use is needed, it must have the owner's signature.

\vskip 6mm
\noindent
The chain separation technology separates data governance and rule governance. The manager is only responsible for rule governance, and all participants are required to upload activity logs in accordance with the rules. Therefore, the intelligent governance model provides a way to clarify the boundaries of each participant's responsibilities, and legitimize the activities of the participants within their own responsibilities.

\subsection{Evaluation method}
\noindent
In order for the entire system to have inherent self-driving force for automatic optimization, each chain needs to be adjusted and optimized based on the results after evaluation. Rewards are provided to the chains with high results and penalties to the chains with poor results, and use economic principles to allow market forces to drive the entire system toward the optimal allocation of resources. There are currently two assessment methods:

\vskip 6mm
\noindent
1. The performance of the chain, the speed of consensus propagation, describe the time from the submission of consensus to the consensus of the whole chain of a transaction.

\vskip 6mm
\noindent
2. The trust degree of the chain describes the trust relationship between the chain and other chains, which can be expressed by a weighted adjacency matrix.

\section{Evaluation calculation}
\label{s:Evaluation calculation}

\subsection{Propagation speed assessment}
\noindent
Due to the complexity of the application scenario and the characteristics of the blockchain system, throughput and performance are inversely proportional to the number of participating consensus nodes. The chain separation technology reduces the number of consensus nodes in a single chain, improves the consensus speed, and reduces the time to reach the consistency of the entire chain. Since the consensus is only within the chain, and other chains are used as read-only nodes to verify data, the data propagation of other chains becomes read-only propagation, the network topology has changed from a mesh structure to a snowflake structure.

\vskip 6mm
\noindent
Due to the large size of the transaction data structure designed by the protocol framework, the block data size is also large. The size of the transmission block data message is related to the transmission delay cost. The smaller the message, the more the additional load. In order to avoid excessive additional load and subsequent additional calculations, a very small inv message is added before each message is transmitted to detect whether a specified block is to be transmitted, and the node that has received. It no longer needs to obtain the block [4].

\vskip 6mm
\noindent
Suppose $ S $ is a node that has not obtained a block, $ I $ is a node that has obtained a block, and the total number of nodes in the chain is $ N $.

\vskip 6mm
\noindent
Definition: $ t $ is the number of transmissions, which is a positive integer.

\vskip 6mm
\noindent
Then the ratios of the two types of nodes to the number of sub-chain nodes at the time of $ t $ are recorded as: $ s(t) $, $ i(t) $, and the number of two types of nodes are: $ S(t) $, $ I(t) $.

\vskip 6mm
\noindent
At the initial moment $ t = 0 $, the initial ratio of the number of acquired nodes to the number of unacquired nodes is $ s_{0} $ and $ i_{0} $.

\vskip 6mm
\noindent
The average number of nodes (the average number of node addresses held by the network) of each node in a propagation cycle is: $ \lambda $.

\vskip 6mm
\noindent
Each propagation marks the nodes that have not obtained blocks in the ratio of $ \lambda*s(t) $ as ``obtained'' ones, so the number of nodes that have obtained blocks is $ N*i(t) $, so there is every day $ \lambda*s(t)*N*i(t) $ nodes to obtain blocks, that is, the number of nodes that have been obtained blocks newly added every day, the differential equation can be obtained:
\noindent
{\large\textbf{
\begin{equation}
N*di(t)/dt=\lambda*s(t)*N*i(t)
\end{equation}
}}

{\large\textbf{
 \begin{equation}
 s(t)+i(t)=1
 \end{equation}
}}

{\large\textbf{
\begin{equation}
\begin{split}
di(t)/dt=\lambda*(1-i(t))*i(t)  \\
\mbox{, and}\; i(0)=i_{0}
\end{split}
\end{equation}
}}

\vskip 6mm
\noindent
Total number of obtained block nodes:
{\large\textbf{
\begin{equation}
I(t)=N*i(t)
\end{equation}
}}

\vskip 6mm
\noindent
Solved according to the natural logarithm:
{\large\textbf{
\begin{equation}
\begin{split}
i\left (t \right )=\frac{i_{0}e^{\lambda t}}{1+i_{0}(e^{\lambda }t+1)}\\
=\frac{1 }{1+(\frac{1}{i_{0}}-1)e^{-\lambda t}}
\end{split}
\end{equation}
}}

\vskip 6mm
\noindent
From the nature of the natural logarithm, it can be obtained that when $ i(t) = 1 $, $ t $ is infinite. Since $ I $, $ S $ and $ t $ are all positive integers, we find that $ t $ makes $ i(t )=N-1/N $, the above formula is transformed into: $ t=1/\lambda ln((N-1)(i_{0}-1)/i_{0}) $

\vskip 6mm
\noindent
Because one block transmission is two api calls (one inv), we may set the average delay of each api call to $ p $,

\vskip 6mm
\noindent
Then the delay of a propagation: $ delay=\lambda*2*p $

\vskip 6mm
\noindent
From $ i(t)=N-1/N $, $ i(t) $ is infinitely close to $1$, but not equal to $1$.

\vskip 6mm
\noindent
Premise: $ N $, $ S $, $ I $, $ t $ are all positive integers

\vskip 6mm
\noindent
Corollary: $ i(t) >= 1 $ when $ t+1 $. Since it is impossible for $ i(t) $ to be $ > 1 $, it can be deduced that $ i(t) = 1 $.

\vskip 6mm
\noindent
Set the final duration: $ T=(t+1)*delay $

\vskip 6mm
\noindent
Calculation result: $ T=2p\;ln((N-1)(i_{0}-1)/i_{0})+2p\lambda $

\subsection{Trust evaluation}

\subsubsection{Trust}
\noindent
The degree of trust refers to the degree to which an entity trusts another entity, where the trust degree of the entity to itself is fixed at 1. The degree of trust can be calculated from the three indicators of credibility, reliability, and intimacy.

\vskip 6mm
\noindent
Suppose: the trust degree of the entity $ i $ to the entity $ j $ is $ Cred_{ij} $ (the degree to which the entity $ j $ trusts the entity $ i $)

{\large\textbf{
\begin{equation}
\left [Cred_{ij} \right ]=\left [NC_{i}\times DPR(G_{i})\times I_{ij}^{\ast} \right]
\end{equation}
}}

\subsubsection{Credibility}
\noindent
Suppose: the credibility is $ NC $, and $ NC^{init} $ is the initial credibility [10]. Reliability is a constantly changing value that is weighted and calculated based on feedback.

\vskip 6mm
\noindent
Note: $ t_{cur} $ and $ t_{pre} $ represent the time gap between the current credibility and the last credibility.

\vskip 6mm
\noindent
$ T(i) $ and $ RP(i) $ indicate the number of weighting calculations and adjustment values.

\vskip 6mm
\noindent
Then, $ NC(i) $ is calculated as follows:
{\large\textbf{
\begin{equation}
\begin{split}
	NC_{i}=NC_{i}^{init}\\ 
+(t_{cur}-t_{pre}+1)\sum_{i}^{j}T(i)\\ 
\ast RP(i)
\end{split}
\end{equation}
}}

\subsubsection{Network reliability}
\noindent
Suppose: entity $ i $ is a distributed network, then the reliability of entity $ i $ is recorded as $ DPR (G_{i}) $ [11].
\newline

{\large\textbf{
\begin{equation}
\begin{split}
DPR(G_{i})=p_{s}[p_{e}p_{v}DPR1(G\ast e)\\
+(1-p_{e}p_{v})DPR1(G-e)]
\end{split}
\end{equation}
}}
\newline

\vskip 6mm
\noindent
Among them, $ p_{s} $ is the probability of normal operation of the starting node $ s $, $ e $ is the network transmission from $ s $ to $ v $, and $ p_{e} $ is the probability of normal operation of the network transmission from $ s $ to $ v $. $ p_{v} $ is the probability of normal operation of node $ v $. $ DPR1(G*e) $ represents the original network $ G $ fusion $ e $ means: the distributed reliability of the merged $ s $ and $ v $, $ DPR1(Ge) $ represents the original network $ G $ after cropping $ e $ distributed reliability.

\subsubsection{Intimacy between entities}
\noindent
The intimacy $ I_{ij} $ between the entity $ i $ and the entity $ j $ is defined by the frequency of messages sent by the two entities [12], and the total number of the entity $ i $ sent to the entity $ j $ is recorded as $ q_{ ij} $, the total number of messages received by the entity is recorded as $ q^{in} $, and the total number of messages sent by the entity is recorded as $ q^{out} $. Then, the formula for calculating intimacy between entities:
{\large\textbf{
\begin{equation}
 I_{ij}^{\ast}=\frac{q_{ij}}{\sqrt{q_{i}^{out}q_{j}^{in}}} 
\end{equation}
\begin{equation}
 q_{i}^{out}=\sum_{j=1}^{n}q_{ij} 
\end{equation}
\begin{equation}
 q_{j}^{in}=\sum_{i=1}^{n}q_{ij} 
\end{equation}
}}

\subsection{Credit result}
\noindent
According to the above formula, the network composed of five sub-chains is calculated and evaluated. The result of the trust degree between them is:

\begin{center}
\textbf{Table 1\quad Result of the trust degree}
\end{center}
\begin{table}[ht]
\begin{tabular}{|p{0.25in}|p{0.4in}|p{0.4in}|p{0.4in}|p{0.4in}|p{0.4in}|} \hline 
Chain & 1 & 2 & 3 & 4 & 5 \\ \hline 
1 & 1 & 0.053785 & 0.164643 & 0.212927 & 0.086383 \\ 
2 & 0.275296 & 1 & 0.068036 & 0.289826 & 0.072169 \\ 
3 & 0.018092 & 0.003233 & 1 & 0.014873 & 0.041154 \\
4 & 0.498808 & 0.157861 & 0.136248 & 1 & 0.158461 \\ 
5 & 0.084898 & 0.114817 & 0.027762 & 0.033741 & 1 \\ \hline 
\end{tabular}
\end{table}

\section{Conclusion}
\noindent
The Internet of Rights(IOR) model is a blockchain application model that tries to solve certain problems faced by the data era with considerations of distributed governance. It formulates governance rules through consensus on roles and permissions, allowing each participant to conduct activities in accordance with the rules, and put the fingerprints of the activity log on the chain.

\vskip 6mm
\noindent
The idea is to reach a consensus on data without knowing what the data is. Therefore data privacy and collaboration no longer conflict. Participants of the chain obtain only the data they need according to their roles and permissions, which reduces data security issues and reduces liability risks, so that the entire link from data collection to use is compliant and legal.

\vskip 6mm
\noindent
The storage consensus of the Internet of Rights(IOR) model makes mutual verification between multiple chains possible and provides a way to resolve the conflict between owners and users. Data owners can specify the scope of users based on roles and permissions, and reasonably divide the rights of owners and users through data authorization and authentication services.

\vskip 6mm
\noindent
The chain separation feature provided by the Internet of Rights(IOR) model separates the role chain, the permission chain and the service chain, and provides a method of separating data governance and rule governance, allowing managers to formulate rules and users to use data according to the rules. It reduces the impact of a single point of failure, improves the availability of the system, reduces the risk of managers that used to manage too much data before, and creates an intelligent work model.

\vskip 6mm
\noindent
The Internet of Rights(IOR) model increases the cost of fraud through immutable and traceable accounting capabilities, retains log records, and can be traced when needed, promotes the relationship of mutual trust between various chains, and enhances the participation in the creation of data value. Therefore, it is necessary to evaluate the participants and make continuous adjustments based on historical records. Participants with higher scores are rewarded, which helps everyone to form a fair and just environment.

\vskip 6mm
\noindent
Under the multi chain consensus mechanism of the Internet of Rights(IOR) system, the UTXO model and the account model will be used together as the two modes of bookkeeping in the future, and the transactions between them can be converted to each other [15]. Multichain consensus integrates dual-ledger inclusive design, which not only adopts the maturity and stability of traditional technology, but also leaves room for new distributed ledger technology, making the two distributed technologies compatible with each other, parallel and complementary [14].

  \end{document}